\documentclass[%
 reprint,
 amsmath,amssymb,
 aps,
prl,
]{revtex4-2}

\usepackage{graphicx}
\usepackage{bm}
\usepackage{amsmath}
\usepackage{hyperref}
\usepackage{esint}
\usepackage{comment,xcolor}
\DeclareMathOperator{\Tr}{Tr}

\DeclareMathOperator{\Real}{Re}

\begin{document}
\title{Quantum limit of precision for phase estimation in squeezing-enhanced interferometry with a single-mode readout}%

\author{Dmitri B. Horoshko}\email{dmitri.horoshko@uni-ulm.de}
\affiliation{Institute for Quantum Optics, Ulm University, D-89081 Ulm, Germany}
\author{Fedor Jelezko}
\affiliation{Institute for Quantum Optics, Ulm University, D-89081 Ulm, Germany}
\date{\today}

\begin{abstract}
We consider an optical interferometer with coherent light in one input and a squeezed vacuum in another. Such an interferometer is known to beat the standard quantum limit of sensitivity to the difference of phase shifts in its arms. We find the ultimate limit of precision for such an interferometer by calculating the quantum Fisher information of the mixed quantum state in one of the interferometer's outputs about the difference phase. We show that, in the vicinity of the black fringe, this information is asymptotically close to the quantum Fisher information about this phase for the two-mode readout. We conclude that the single-mode readout is optimal for phase estimation in squeezing-enhanced interferometry and allows for the Heisenberg scaling of precision. We also show that the optimal local measurement in the vicinity of the black fringe consists of amplifying the output field in a phase-sensitive way and measuring its photon number.
\end{abstract}
\maketitle

\textit{Introduction}. Optical interferometers, from the evidence of the wave nature of light in Young's double-slit experiment to the refutation of the aether hypothesis by Michelson and Morley \cite{Born&Wolf}, and to the observation of gravitational waves by the LIGO and VIRGO consortia \cite{Abbott16}, were always key instruments for exploring the laws of nature. Today, they are used in a wide variety of applications, including gravitational astronomy \cite{Abbott16},  high angular resolution astronomy  \cite{Eisenhauer23}, optical chemical sensors \cite{McDonagh08}, optical coherence tomography \cite{DeBoer17}, and biological sensing \cite{Taylor13}. The idea of illuminating the unused input port of the Michelson interferometer with squeezed light \cite{Caves81} proved to be extremely fruitful and led to a significant increase in the sensitivity of modern gravitational wave detectors \cite{Hild09,Ligo11,Aasi13,Tse19,Ganapathy23}. An extension of this idea led to the construction of nonlinear optical interferometers in which the squeezed light is generated internally \cite{Yurke86,Caves20,Salykina23}. Realization of such interferometers with the help of high-gain parametric down-conversion enabled Fourier-transform infrared spectroscopy \cite{Hashimoto24} and low-coherence interferometry \cite{Zotti25} in the mid-infrared spectral region.

The ultimate limit of squeezing-enhanced interferometry, optimized over all possible measurements, is given by the quantum Fisher information (QFI) \cite{Helstrom-book} of the output light about the phase difference between the interferometer arms. It was found that in the case of two-mode readout (both interferometer outputs are measured), the QFI is given by $\mathcal{F}_0=|\alpha|^2e^{2r}+\sinh^2r$, where $\alpha$ is the amplitude of the input coherent state and $r$ is the squeezing parameter of the input squeezed vacuum \cite{Jarzyna12}. This formula gives the scaling $\mathcal{F}_0\sim\langle \hat N_+\rangle^2$, where $\langle \hat N_+\rangle$ is the mean total number of photons in the interferometer, which is the fastest possible scaling known as the Heisenberg limit of sensitivity \cite{Giovannetti04}. It was also found that in the scenario where the number of photons is measured in both interferometer output modes, the classical Fisher information about the difference phase is also given by $\mathcal{F}_0$ \cite{Pezze08,Lang13}, which means that the measurement of photon number in two modes is optimal.

However, a two-mode readout is often difficult to realize in practice, and, for example, it is considered unfeasible in gravitational wave interferometry for a number of technical reasons \cite{Hild09,Ligo11,Aasi13,Tse19,Ganapathy23}. In this paper, we analyze a single-mode readout of a squeezing-enhanced Mach-Zehnder interferometer (MZI) and calculate the single-mode QFI on the difference phase. The state of a single output mode is generally mixed, which complicates the calculation of QFI, in contrast to the two-mode readout, where the state is pure. However, it was recently shown that the QFI has a compact expression for a Gaussian state with a known Williamson decomposition \cite{Safranek19}. Since the single-mode state is Gaussian and its Williamson decomposition can be readily obtained, the calculation of QFI is straightforward. We show that the single-mode QFI about the difference phase is also given by $\mathcal{F}_0$, which means that the single-mode readout is optimal and provides the Heisenberg scaling in the absence of loss.

\begin{figure}[ht!]
\centering
\includegraphics[width=\linewidth]{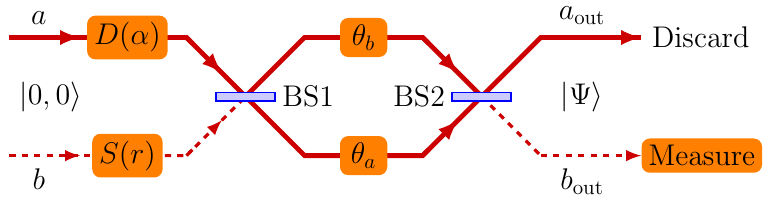}
\caption{Mach-Zehnder interferometer created by beam splitters BS1 and BS2. The interferometer has a coherent (mode $a$) and a squeezed (mode $b$) fields at two its inputs. Unknown phase shifts $\theta_a$ and $\theta_b$ are introduced inside the interferometer to modes $a$ and $b$ respectively. The task is to infer the difference phase $\theta=\theta_a-\theta_b$ from the measurement data of mode $b_\text{out}$ alone.
\label{fig:MZI}}
\end{figure}

\textit{Field transformation in MZI}. We consider an MZI with coherent light in one input and a squeezed vacuum in another, as shown in Fig.~\ref{fig:MZI}, which is equivalent to the Michelson interferometer considered by Caves \cite{Caves81}.  

Two modes $a$ and $b$, initially in the vacuum state of both modes $|0,0\rangle$, pass through the laser (mode $a$) and the squeezer (mode $b$). The effect of laser generation is described by the displacement operator $\hat D=\exp\left(\alpha a^\dagger-\alpha a\right)$, where $a$ is the photon annihilation operator for this mode, $a^\dagger$ is its Hermite conjugate and $\alpha$ is the complex field amplitude \cite{LoudonBook}. The squeezer, typically realized by a degenerate optical parametric amplifier, reduces the standard deviation of a mode quadrature by the factor $e^{-r}$, where $r>0$ is the squeezing parameter; this action is described by the squeezing operator $\hat S=\exp\left[\frac{r}2(b^2-b^{\dagger 2})\right]$, where $b$ is the photon annihilation operator for this mode \cite{LoudonBook}. Subsequently, the two fields are mixed on a balanced beam splitter BS1, whose action is described by the beam-splitter operator $\hat T=\exp[\frac{\pi}4(a^\dagger b-ab^\dagger)]$. Inside the interferometer, the modes $a$ and $b$ experience phase shifts $\theta_a$ and $\theta_b$, respectively; this action is described by the phase shift operator $e^{i\hat G}$ where $\hat G= \theta_a a^\dagger a+\theta_b b^\dagger b$ is the generator of the phase shifts. The modes are then recombined on the beam splitter BS2, characterized by the evolution operator $\hat T^\dagger$. The state of both modes at the interferometer output is
\begin{equation}\label{Psi}
\left|\Psi\right\rangle = \hat T^\dagger e^{i\hat G}\hat T\hat S\hat D|0,0\rangle \equiv \hat U|0,0\rangle,    
\end{equation}
where we have introduced the operator of the entire evolution $\hat U$.

As indicated above, we are interested in the case where only the mode $b_\text{out}$ is measured, while the mode $a_\text{out}$ is discarded. In the Heisenberg picture, the photon annihilation operator of the output field of mode $b$ is $b_\text{out} = \hat U^\dagger b \hat U$ and it can be expressed via the input vacuum-field operators as 
\begin{eqnarray}\label{bout}
b_\text{out} &=& e^{i\Phi/2}\left[ c(b\cosh r-b^\dagger\sinh r) +is(a+\alpha) \right],    
\end{eqnarray}
where we have introduced the sum and difference phases, $\Phi=\theta_a+\theta_b$ and $\theta=\theta_a-\theta_b$, respectively, and the shortcuts $c=\cos(\theta/2)$ and $s=\sin(\theta/2)$.

\textit{$N$-precision}. The photon number operator for the  mode $b_\text{out}$ in the interferometer output is $\hat N_b= b_\text{out}^\dagger b_\text{out}$. Its mean and variance are obtained from Eq.~(\ref{bout}) as
\begin{eqnarray}\label{Nb}
\langle\hat N_b\rangle &=& c^2\sinh^2r + s^2|\alpha|^2\\\nonumber
\langle\Delta\hat N_b^2\rangle &=& s^4|\alpha|^2 + s^2c^2\sinh^2 r +2c^4\cosh^2 r\sinh^2 r\\\label{DeltaN} 
&+&s^2c^2\left|\alpha\cosh r+\alpha^*\sinh r\right|^2,
\end{eqnarray}
where $\Delta\hat N_b=\hat N_b-\langle\hat N_b\rangle$ and the angular brackets stand for the averaging with the vacuum field. We see from Eq.~(\ref{DeltaN}) that the variance is determined by the phase of $\alpha$. The minimal photon number variance is reached for a purely imaginary $\alpha=-i|\alpha|$, which will be used in the following.

We observe that the mean photon number depends only on $\theta$, but not on $\Phi$. Measurement of $\langle\hat N_b\rangle$, therefore, allows one to measure $\theta$. The precision of this measurement is given by the inverse of the parameter variance $P_\theta=1/\langle\Delta\theta^2\rangle$ \cite{Bernardo-Smith,Horoshko25} and is obtained from the error propagation rule as
\begin{eqnarray}\label{Ptheta}
P_\theta &=& \left(\frac{\partial\langle\hat N_b\rangle}{\partial\theta}\right)^2 \frac1{\langle\Delta\hat N_b^2\rangle}\\\nonumber
&=& |\alpha|^2\frac{(1-\epsilon)^2\sin^2(\theta/2)\cos^2(\theta/2)}
{\sin^4(\theta/2) + e^{-2r}\sin^2(\theta/2)\cos^2(\theta/2)+\epsilon g},   
\end{eqnarray}
where $\epsilon=\sinh^2r/|\alpha|^2$ is the ratio of mean number of photons in the squeezed and coherent modes and $g=\sin^2(\theta/2) \cos^2(\theta/2) + 2\cosh^2r\cos^4(\theta/2)$. The  maximum precision is reached at the optimal difference phase, where  
\begin{eqnarray}
\theta_\text{opt} &=&\pm2\arctan\left(\sqrt{\sinh2r/\sqrt{2}|\alpha|}\right),\\\label{Popt}
P_{\theta_\text{opt}} &=& |\alpha|^2e^{2r} \frac{(1-\epsilon)^2}
{1 + e^{2r}\left(\epsilon+2\sqrt{2\epsilon\cosh^2r}\right)} . 
\end{eqnarray}

\begin{figure}[ht!]
\centering
\includegraphics[width=\linewidth]{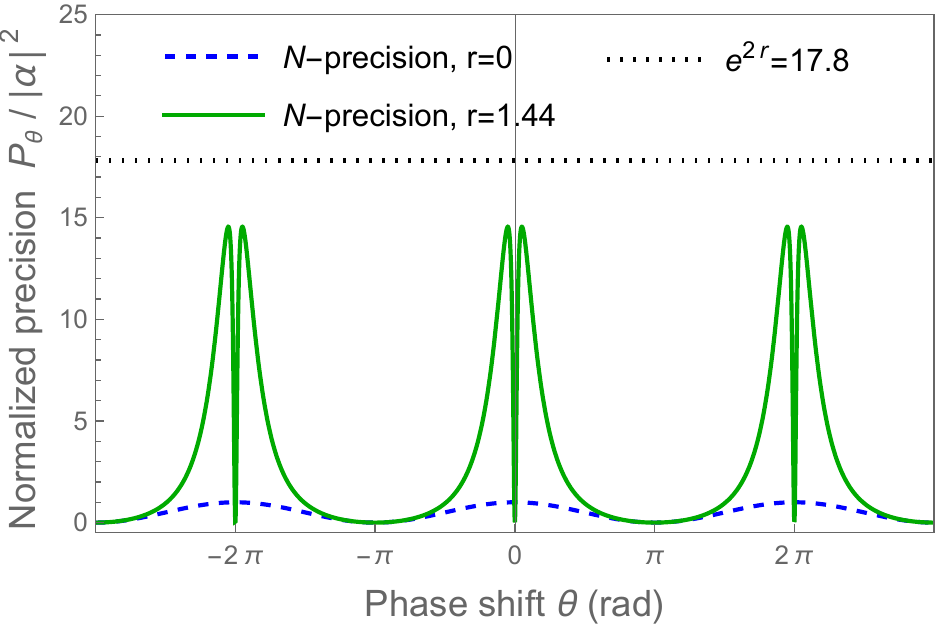}
\caption{Normalized $N$-precision for measuring the difference phase $\theta$ by detecting the number of photons in mode $b$ in the strong-field regime (the mode $a$ has $|\alpha|^2=10^6$ photons at the input). The standard quantum limit is shown by the blue dashed line. The solid green line corresponds to the routinely obtained squeezing $20r\lg e=12.5$ dB and shows that the normalized precision drops to zero at the black fringe ($\theta=0$) but raises to about 15 in its vicinity. The maximal value of $e^{2r}$ corresponds to the limit $|\alpha|^2\to\infty$ (black dotted line).
\label{fig:Nprecision}}
\end{figure}

In strong-field applications of squeezing-enhanced interferometry, such as gravitational wave astronomy, the mean photon number in the coherent mode, $|\alpha|^2$, can be very hight. Indeed, during the first implementation of squeezing enhancement in GEO 600 detector \cite{Ligo11}, the laser power was $\mathcal{P}=12$ W and the coherence time of the squeezer was $\tau_c=10^{-4}$ s, which gives the number of photons in the coherence time $n_c=\mathcal{P}\lambda_L\tau_c/(hc_L)=6.4\times10^{15}$, where $h$ is the Planck's constant, $c_L$ the speed of light in vacuum, and $\lambda_L=1.064$ $\mu$m the wavelength of the laser light.  The value of $n_c$ is roughly equal to $|\alpha|^2$ of the theoretical model up to optical loss. At the same time, the squeezing factor $e^{2r}$ is limited by the level of squeezing in the degenerate optical parametric amplifier and does not exceed $10^2$, in view of the highest up-to-date measured squeezing $20r\lg e=15$ dB \cite{Vahlbruch16}. In such applications, $\epsilon$ is very close to zero, and, in the vicinity of the ``black fringe'' $\theta=0$, the precision $P_\theta$ experiences a rapid jump from $P_0=0$ to $P_{\theta_\text{opt}}\approx|\alpha|^2e^{2r}$ \cite{Caves81}, as illustrated in  Fig.~\ref{fig:Nprecision}. 

\textit{Gaussian state of mode $b$}. The ultimate limit of precision for inferring the phases $\Phi$ and $\theta$ from the measurements applied to the mode $b$ at the interferometer output is given by the quantum Fisher information matrix (QFIM) \cite{Helstrom-book}. The state of this mode belongs to the class of Gaussian single-mode states, for which the general algorithm for finding the QFIM is well established \cite{Pinel13,Gao14} and is relatively simple when the Williamson decomposition is known \cite{Safranek19,Safranek15}. Let us find this decomposition. 

This Gaussian state of mode $b$ is fully determined by the mean field
\begin{equation}\label{d}
d\equiv\langle b_\text{out}\rangle =|\alpha|e^{i\Phi/2}\sin(\theta/2)
\end{equation}
and the complex covariance matrix  \cite{Safranek19} 
\begin{equation}\label{sigma}
\boldsymbol\sigma = \left(
    \begin{array}{cc}
        1+2C_N & 2C_A \\
        2C_A^* & 1+2C_N
    \end{array}
    \right),
\end{equation}
where $C_N$ and $C_A$ are the normal and anomalous second-order central moments of the field, respectively, defined as 
\begin{eqnarray}\label{CN}
C_N &\equiv& \langle\Delta b_\text{out}^\dagger\Delta b_\text{out}\rangle = \frac12(1+\cos\theta)\sinh^2r,\\\label{CA}
C_A &\equiv& \langle\Delta b_\text{out}^2\rangle = -\frac12e^{i\Phi}(1+\cos\theta)\cosh r\sinh r,
\end{eqnarray}
where $\Delta b_\text{out} =b_\text{out}-d$. The Williamson decomposition of this single-mode covariance matrix is $\boldsymbol\sigma=\lambda\mathbf{S} \mathbf{S}^\dagger$, where  $\mathbf{S}$ is a symplectic matrix, while $\lambda=\sqrt{\det \boldsymbol\sigma}$ is a positive number known as the symplectic eigenvalue and characterizing the state purity: $\lambda=1$ for a pure state and $\lambda>1$ for a mixed state. We find from Eqs.~(\ref{sigma}) -- (\ref{CA}) that  
\begin{eqnarray}\label{lambda}
\lambda &=& \sqrt{1+\sin^2\theta\sinh^2r}, \\\label{S}
\mathbf{S} &=& \left(
\begin{array}{cc}
\cosh r_\text{out}  & -e^{i\Phi}\sinh r_\text{out} \\
-e^{-i\Phi}\sinh r_\text{out}        & \cosh r_\text{out}
\end{array}
\right),
\end{eqnarray}
where 
\begin{equation}
r_\text{out} = \frac12\ln\frac{\sin^2\frac\theta2 + e^{2r}\cos^2\frac\theta2}\lambda \end{equation}
is the output squeezing parameter. Note, that the output state of mode $b_\text{out}$ is a displaced squeezed thermal state and the variances of its squeezed and stretched quadratures are $e^{2r_\lambda\pm 2r_\text{out}}$, where $r_\lambda=\frac12\ln\lambda$.

Equation (\ref{lambda}) shows that the state of the mode $b$ is pure ($\lambda=1$) at $\theta=0$ and $\theta=\pi$. As follows from Eq.~(\ref{Psi}), this state is pure squeezed at the black fringe $\theta=0$, where $r_\text{out}=r$. At the white fringe, $\theta=\pi$, the state is pure coherent with $r_\text{out}=0$, while it is mixed ($\lambda>1$) at all other values of $\theta$. 

\textit{QFIM for single-mode readout}. In the intervals between the black and white fringes, the QFIM of mode $b$ about parameters $(\Phi,\theta)$ has a relatively simple expression \cite{Safranek19}
\begin{equation}\label{Qmixed}
Q^{kl}_\theta=\frac{4\lambda^2}{\lambda^2+1} \Real\left\{J_k^*J_l\right\} +\frac{\partial_k\lambda\partial_l\lambda}{\lambda^2-1} +2\partial_k\vec{d}^\dagger\boldsymbol\sigma^{-1}\partial_l\vec{d},
\end{equation}
where the matrix indices $k$ and $l$ take values from the parameter vector $(\Phi\,\,\theta)$, $\partial_k$ means a derivative with respect to the parameter $k$, $\vec{d}=(d\,\, d^*)^T$, and the elements of the vector $(J_\Phi\,\, J_\theta)$ are expressed via the elements of the symplectic matrix $\mathbf{S}$ as
\begin{equation}
J_k=S_{11}^*\partial_kS_{12}-S_{12}\partial_kS_{11}^*.  
\end{equation}
Note that the QFIM generally depends on both parameters, but we indicate only its dependence on $\theta$ explicitly, because only this parameter determines $\lambda$ and therefore affects the expression for the QFIM.

At the black and white fringes, where the state of the mode becomes pure, QFIM may be discontinuous if the second derivative of the symplectic eigenvalue $\lambda$ is nonzero \cite{Safranek19,Seveso20}. We find from Eq.~(\ref{lambda}), that, at $\theta=0$ or $\theta=\pi$, $\partial_\theta^2\lambda=\sinh^2r$, i.e., it is nonzero when $r>0$. In the following, we will be interested in the vicinity of the black fringe, where the maximal precision is reached for squeezing-enhanced interferometry. We arrive at the conclusion that the QFIM experiences a jump from the value ${\bf Q}_0$ at $\theta=0$ to ${\bf Q}_{0+}$ at $\theta\to0$. The elements of these matrices are \cite{Safranek19}
\begin{eqnarray}\label{Q0}
Q_0^{kl} &=& \frac14\Tr\left\{ \boldsymbol\sigma^{-1} \partial_k\boldsymbol\sigma\boldsymbol\sigma^{-1} \partial_l\boldsymbol\sigma\right\}+2\partial_k\vec{d}^\dagger\boldsymbol\sigma^{-1}\partial_l\vec{d},\\\label{Q0+}
Q_{0+}^{kl} &=& \frac14\Tr\left\{2\boldsymbol\sigma^{-1} \partial_k\partial_l\boldsymbol\sigma-\boldsymbol\sigma^{-1} \partial_k\boldsymbol\sigma\boldsymbol\sigma^{-1} \partial_l\boldsymbol\sigma\right\}\\\nonumber
&+& 2\partial_k\vec{d}^\dagger\boldsymbol\sigma^{-1}\partial_l\vec{d},
\end{eqnarray}
where all derivatives are evaluated at $\theta=0$.

\begin{figure}[ht!]
\centering
\includegraphics[width=\linewidth]{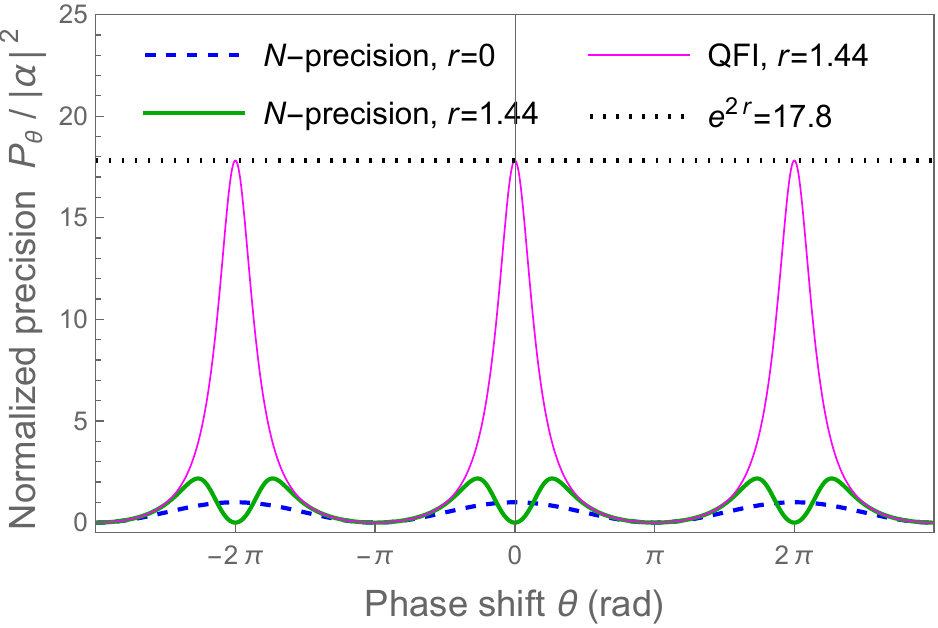}
\caption{Normalized $N$-precision and normalized QFI $Q^{\theta\theta}_\theta/|\alpha|^2$ in the weak-field regime, where the mode $a$ has $|\alpha|^2=10^3$ photons at the input. The standard quantum limit is obtained when no squeezing is used (blue dashed line). The QFI (thin magenta line) is much higher than the $N$-precision (thick green line) in the vicinity of the black fringe and even exceeds the limiting value of the latter $e^{2r}$ (black dotted line).
\label{fig:QFI}}
\end{figure}

Substituting Eqs.~(\ref{sigma}), (\ref{lambda}) and (\ref{S}) into Eqs.~(\ref{Qmixed}), (\ref{Q0}) and (\ref{Q0+}), we obtain $Q^{\Phi\theta}_\theta=0$ for all values of $\theta$, which means that the limits of precision for measuring $\Phi$ and $\theta$ are given by the corresponding matrix elements of QFIM, similar to the two-mode readout \cite{Jarzyna12,Demkowicz15}. 
For this reason, in what follows, we concentrate on the precision limit in measuring the difference phase $\theta$ given by the QFIM element $Q^{\theta\theta}_\theta$. 
We also find from Eqs.~(\ref{Q0}) and (\ref{Q0+}) that 
\begin{equation}\label{QTheta}
Q_{0}^{\theta\theta}=|\alpha|^2e^{2r}, \quad Q_{0+}^{\theta\theta}= |\alpha|^2e^{2r} +\sinh^2r.  
\end{equation}
On the one hand, this means that, in the vicinity of the black fringe, the QFI is higher than the $N$-precision given by Eq.~(\ref{Popt}). Therefore, an optimal measurement combined with statistical inference by maximum-likelihood or Bayesian estimation would provide a better estimate of the difference phase than detection of the photon number. The difference between the $N$-precision and the ultimate quantum limit given by the QFI is especially large in the weak-field regime, where $|\alpha|^2$ is relatively small (see Fig.~\ref{fig:QFI}), which may be required for biomedical applications. 

Asymptotically at $\langle \hat N_+\rangle\to\infty$, the QFI behaves as $Q^{\theta\theta}_{0+} \sim \langle \hat N_+\rangle^2$, where $\langle \hat N_+\rangle$ is the mean total number of photons in the interferometer, i.e. in the Heisenberg limit of precision \cite{Giovannetti04,Demkowicz15,Paris95}. In contrast, the $N$-precision behaves as $P_\theta\sim\langle \hat N_+\rangle^\frac32$, at a much slower pace \cite{Demkowicz15}. 

On the other hand, we see that $Q_{0+}^{\theta\theta} = \mathcal{F}_0$, where $\mathcal{F}_0$ is the QFI for the two-mode readout, as mentioned in the Introduction. Therefore, we conclude that the single-mode readout is optimal for estimating the difference phase. 

\textit{Optimal measurement}. The optimal measurement of mode $b_\text{out}$, if it exists, is a projection on the joint eigenbasis of the two symmetric logarithmic derivatives (SLD) \cite{Safranek19}
\begin{eqnarray}
\mathcal{L}^k_\theta &=& \Delta\vec B^\dagger (\mathbf{S}^{-1})^\dagger\mathbf{W}^k \mathbf{S}^{-1}\Delta\vec B - \frac{\lambda\partial_k\lambda}{\lambda^2-1} \\\nonumber 
&+& 2\Delta\vec B^\dagger \boldsymbol\sigma^{-1}\partial_k\vec{d},
\end{eqnarray}
where the upper index of $\mathcal{L}^k_\theta$ refers to the parameter over which the derivative is taken ($\Phi$ or $\theta$), while the lower index gives the value of $\theta$, $\Delta\vec B = \vec B - \vec d$, $\vec B = (b_\text{out}\,\, b_\text{out}^\dagger)^T$, and the  Hermitian matrix $\mathbf{W}^k$ has the matrix elements 
\begin{eqnarray}
W^k_{11} &=& W^k_{22} = \frac{\partial_k\lambda}{\lambda^2-1},\\
W^k_{12} &=& W^{k*}_{21} = \frac{2\lambda}{\lambda^2+1}J_k.
\end{eqnarray}
Noticing that $\boldsymbol\sigma^{-1}=\lambda^{-1}(\mathbf{S}^{-1})^\dagger \mathbf{S}^{-1}$ and substituting the values of $d$, $\lambda$ and $\mathbf{S}$ from Eqs.~(\ref{d}), (\ref{lambda}), and (\ref{S}), we obtain the following SLDs in the vicinity of the dark fringe
\begin{eqnarray}\label{LPhi}
\mathcal{L}^\Phi_{0+} &=& i\cosh{r}\sinh{r} \left(\tilde b_\text{out}^2 e^{-i\Phi} - \tilde b_\text{out}^{\dagger2} e^{i\Phi}\right),\\\label{Ltheta}
\mathcal{L}^\theta_{0+} &=& 2\cot{\theta} \tilde b_\text{out}^{\dagger} 
\tilde b_\text{out},
\end{eqnarray}
where 
\begin{equation}\label{tildebout}
\tilde b_\text{out} = b_\text{out}\cosh{r} +  b_\text{out}^\dagger e^{i\Phi}\sinh{r}. 
\end{equation}

The SLD related to $\theta$, Eq.~(\ref{Ltheta}), is proportional to $\cot{\theta}$, which tends to infinity as $\theta$ approaches 0. In practice, $\theta$ has a small nonzero value, so $\cot\theta$ is finite. Calculating the commutator of the two SLDs, we find
\begin{equation}
\lim_{\theta\to0}\left\langle 
\left[\mathcal{L}^\Phi_{\theta},\mathcal{L}^\theta_{\theta}\right]
\right\rangle = 0,    
\end{equation}
which means that both parameters can be estimated at the quantum Cram\'er-Rao bound asymptotically at a high number of trials, where a collective measurement of multiple probes may be required in general \cite{Ragy16}. However, we are interested in estimating the difference phase $\theta$ only and therefore can treat the sum phase $\Phi$ as a nuisance parameter. This allows us to find a measurement that saturates the quantum Cram\'er-Rao bound locally for a finite number of measurements, as shown below.

\textit{Locally optimal readout}. 
Choosing the working point very close to the dark fringe at $\Phi=0$, $\theta=\theta_0\ll1$, we observe from Eq.~(\ref{Ltheta}) that the optimal measurement corresponds to the eigenbasis of operator $\tilde b_\text{out}^{\dagger} 
\tilde b_\text{out}$. As Eq.~(\ref{tildebout}) shows, $\tilde b_\text{out}$ is the photon annihilation operator for the output mode $b_\text{out}$ upon additional squeezing with parameter $r$ at angle $\Phi+\pi$. The optimal measurement basis thus depends on the unknown parameter $\Phi$. However, for the purpose of a \emph{local} estimation, the eigenbasis of SLD can be fixed to the working point. 

Indeed, write the eigenvalues of  $\mathcal{L}^\theta_\theta$ as $\ell_n^{\Phi,\theta}$ and the corresponding eigenvectors as $|\ell_n^{\Phi,\theta}\rangle$, all of them depending on $\Phi$ and $\theta$ in general. Measuring the field with the density operator $\rho_{\Phi,\theta}$ in the working-point basis $|\ell_n^{0,\theta_0}\rangle$, we obtain the probability of outcome $n$ as $p_n(\Phi,\theta)=\langle \ell_n^{0,\theta_0}|\rho_{\Phi,\theta}|\ell_n^{0,\theta_0}\rangle$ with $\partial_\theta p_n= \langle \ell_n^{0,\theta_0}|\partial_\theta\rho_{\Phi,\theta}|\ell_n^{0,\theta_0}\rangle = \ell_n^{0,\theta_0} p_n$, where we have used the defining equation of SLD, $\partial_\theta \rho_{\Phi,\theta} = (\mathcal{L}^\theta_\theta\rho_{\Phi,\theta} + \rho_{\Phi,\theta}\mathcal{L}^\theta_\theta)/2$ \cite{Helstrom-book}. The corresponding Fisher information on the difference phase $\theta$ is
\begin{equation}
F_{\theta_0}^{\theta\theta} = \sum_n\frac{(\partial_\theta p_n)^2}{p_n} = \sum_n \left(\ell_n^{0,\theta_0}\right)^2p_n = \langle(\mathcal{L}^\theta_{\theta_0})^2\rangle,
\end{equation}
which is the definition of $Q_{\theta_0}^{\theta\theta}$ \cite{Helstrom-book}. The local SLD basis thus represents the locally optimal measurement. 

\begin{figure}[ht!]
\centering
\includegraphics[width=\linewidth]{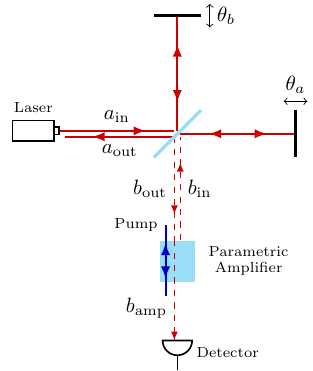}
\caption{Michelson interferometer equivalent to Mach-Zehnder interferometer of Fig.~\ref{fig:MZI} with the optimal single-mode readout in the vicinity of the dark fringe. The same degenerate optical parametric amplifier is used for squeezed state preparation and for readout.
\label{fig:Michelson-readout}}
\end{figure}

Measurement in the eigenbasis of operator $\tilde b_\text{out}^{\dagger} 
\tilde b_\text{out}$ at $\Phi=0$ is carried out by passing the mode $b_\text{out}$ through a squeezer (optical parametric amplifier) with the squeezing parameter $-r$ and measuring the number of photons in the amplified mode $b_\text{amp}$. For this purpose, the same squeezer can be used as for the input state in the reverse direction with the pump phase shifted by $\pi$ with respect to the signal. The scheme of a Michelson interferometer with this kind of readout is shown in Fig.~\ref{fig:Michelson-readout}.

We write the photon annihilation operator of the output mode of the squeezer (in the second passage) as $b_\text{amp} = b_\text{out}\cosh r + b_\text{out}^\dagger\sinh r$. Note that the quadrature that was squeezed in the first passage is stretched in the second. Defining the number of photons in this amplified mode as $\hat N_\text{amp}=b_\text{amp}^\dagger b_\text{amp}$, we calculate the corresponding $N$-precision
\begin{equation}\label{Pthetaamp}
P_\theta^\text{amp} = \left(\frac{\partial\langle\hat N_\text{amp}\rangle}{\partial\theta}\right)^2 \frac1{\langle\Delta\hat N_\text{amp}^2\rangle} = |\alpha|^2e^{2r} +\sinh^2r.   
\end{equation}
This value coincides with the value of $Q_{0+}^{\theta\theta}$ given by Eq.~(\ref{QTheta}), which means that not only the measurement is optimal, but also the estimation of $\theta$ from the measured value of $N_\text{amp}$ can be made for every shot by inverting the dependence of $\langle\hat N_\text{amp}\rangle$ on $\theta$ and this estimation saturates the Cram\'er-Rao bound. $P_\theta^\text{amp}$ also coincides with the QFI about $\theta$ for the two-mode readout, $\mathcal{F}_0$, which confirms that the described type of readout is optimal for measuring the difference phase in squeezing-enhanced interferometry. It should be stressed that the Fock-state projection of the two output interferometer modes is optimal globally, that is, for any values of $\Phi$ and $\theta$ \cite{Pezze08}, while the single-mode readout is optimal only locally in the vicinity of the dark fringe, at $\Phi=0$, $\theta\ll1$.  

We also note that a similar type of readout was considered by Caves in the paper in which squeezing-enhanced interferometry was proposed for the first time \cite{Caves81}. However, Caves considered amplification of the mode $b_\text{out}$ as a means to counter the sensitivity drop caused by the non-unit quantum efficiency of the photodetector and proposed to use the output mode $a_\text{out}$ carrying the sum phase $\Phi$ to realize the squeezing at the angle $\Phi+\pi$. Our study shows that this complicated technique is not necessary for a local estimation of $\theta$ and the phase-sensitive amplification can be done at the angle of squeezing $\pi$ independently of the small variations of the sum phase.

\textit{Conclusions}. We have shown that the single-mode readout is optimal for the estimation of the difference phase, i.e., it provides the same ultimate precision as the two-mode readout. This means that when building a squeezing-enhanced interferometer and aiming for the Heisenberg precision scaling, an optimal measurement for one output mode can be realized, which is much simpler than the measurement of both output modes. We note that a similar task of discarding one of the two modes was recently considered for nonlinear interferometers in Yurke and Mandel configurations \cite{Kranias25}, but the QFI for single-mode readout was always lower than for the two-mode readout. In contrast, a linear squeezing-enhanced interferometer demonstrates equal level of QFI for both types of readout.

One area of application of this fundamental result is gravitational astronomy  \cite{Hild09,Ligo11,Aasi13,Tse19,Ganapathy23} where single-mode readout is the only option since the other output of the Michelson interferometer is very noisy due to the high back reflection of the powerful laser light used in modern gravitational wave detectors. Another area of application is biological sensing \cite{Taylor13,Taylor16} where the weak-field regime is used and the improvement of sensitivity predicted for $P_\theta^\text{amp}$ may be significant compared to $P_\theta$.

\textit{Acknowledgments}. The authors thank Liam McGuinness and Ilya Karuseichyk for enlightening discussions. This work was supported by QuantERA II Programme (project EXTRASENS). It was funded by Horizon 2020 Framework Programme (European Union) via grant 101017733 and Bundesministerium für Forschung, Technologie und Raumfahrt (Germany) via grant 13N16935.

\bibliography{Bib-Interferometry2025.bib}
\end{document}